\documentclass[a4paper]{revtex4}
\usepackage{graphicx}
\usepackage{amsmath}
\usepackage{epsfig}
\usepackage{color}

\begin{document}

\title{Nanomechanics of a Hydrogen Molecule Suspended between Two Equally Charged Tips}
\author{W. Schattke $^{1,2}$, T. Frederiksen $^{1,5}$, M.A. Van Hove $^{3}$ and  R.~D\'{\i}ez Mui\~{n}o$^{1,4}$}
\affiliation{$^{1}$ Donostia International Physics Center DIPC, P. Manuel de Lardizabal 4,
20018 San Sebasti\'an, Spain}
\affiliation{$^{2}$ Institut f\"ur Theoretische Physik und Astrophysik der Christian-Albrechts-Universit\"at,
Leibnizstra{\ss}e 15, 24118 Kiel, Germany}
\affiliation{$^{3}$ Department of Physics and Institute of Computational and Theoretical Studies (ICTS), Hong Kong Baptist University,
224 Waterloo Road, Kowloon, Hong Kong}
\affiliation{$^{4}$ Centro de F\'{\i}sica de Materiales CFM (CSIC-UPV/EHU) -
Materials Physics Center MPC, P. Manuel de
Lardizabal 5, 20018 San Sebasti\'an, Spain}
\affiliation{$^{5}$ IKERBASQUE, Basque Foundation for Science, E-48011, Bilbao, Spain}

%
%
%

\begin{abstract}
Geometric configuration and energy of a hydrogen molecule centered between two point-shaped tips of equal charge are calculated with the
variational quantum Monte-Carlo (QMC) method without the restriction of the Born-Oppenheimer (BO) approximation. Ground state
nuclear distribution, stability, and low vibrational excitation are investigated. Ground state results predict significant
deviations from the BO treatment that is based on a potential energy surface (PES) obtained with the same QMC accuracy. The quantum mechanical
distribution of molecular axis direction and bond length at a sub-nanometer level is fundamental for understanding nanomechanical dynamics
with embedded hydrogen. Because of the tips' arrangement, cylindrical symmetry yields a uniform azimuthal distribution of the molecular
axis vector relative to the tip-tip axis. With approaching tips towards each other, the QMC sampling shows an increasing loss of spherical symmetry
with the molecular axis still uniformly distributed over the azimuthal angle but peaked at the tip-tip direction for negative tip charge while peaked at the equatorial plane for positive charge. This directional behavior can be switched between both stable configurations by changing the sign of the
tip charge and by controlling the tip-tip distance. This suggests an application in the field of molecular machines.
\end{abstract}



\maketitle
\section{Introduction}
Growing interest in nanomechnical properties is reflected by work spreading from the chemistry of synthesizing suitable
nanoscale molecules to the physics of manipulating those objects isolated or adsorbed on surfaces and nanostructures.
The "Supramolecular chemistry"~\cite{Lehn} dates back to the work on
molecular switches through the synthesis of catenanes and rotaxanes~\cite{Stoddart} leading to nanoscale machines such as artificial
molecular pumps~\cite{Chuyang}, molecular wheels, or even nanocars.~\cite{Joachim,Kudernac,Rapenne}.

In addition to reaction chemistry methods, an external physical tool can be used for manipulation, i.e. an interaction with the chemical
object via electromagnetic fields. There is a vast literature on these topics from where we cite only three examples
which come closer to our investigations. Nanocavities built by atomic assembly
at a metal surface~\cite{HuiWang, Carbonell, Lotze} as well as two
tunneling tips forming a bridge~\cite{Smit, Djukic, Halbritter, Trouwborst} are both able to hold in place,  bind and subsequently excite a molecule to perform the desired motion,
e.g. vibration or rotation. There, the variations in the second derivative of the current-voltage characteristic and the drop
of the inelastic current occurring at the same specific voltage allow interaction with the vibrations. These examples show that
a captive hydrogen molecule, which can serve as the basis of complex nanomechanics, can be accurately accessed by experiment and
ab-initio theory.

In the light of these developments we investigate a system which combines simplicity at an ab-initio level with feasibility
at the nanoscale: a hydrogen molecule bridging two STM-like tips, each with an equal point charge, that  enable external control of the
molecule by changing charge and distance. Fundamental work on this molecule ranges from high-precision calculations~\cite{Kolos,Pachucki}
to high-precision measurement, as e.g. highly accurate vibrational determinations~\cite{Dickenson}.
It extends as well into the challenge of examining
the generally used Born-Oppenheimer assumption~\cite{Yarkony, Sutcliffe, Tubman}. The latter is
relevant to this work because the light hydrogen molecule is highly sensitive to non-adiabatic corrections, i.e. the quantum mechanical
nuclear motion coherently taken into account within the QMC electronic calculation.

The ground state energy and the first vibrational excitation are here chosen as the main quantities to probe the reaction of the hydrogen
nuclear motion upon the action of an external static electric field. This system has more general importance besides the two mentioned
examples and their detailed realizations, since the molecule can be detached from the STM tips and placed in another stable electrostatic field
environment produced by a molecular neighborhood of similar symmetry which may allow studying processes  underlying reaction chemistry.

The rotational degrees of freedom of the free molecule in unperturbed space will be restricted by an external perturbation and so will the zero
point energy as well. This will influence the interaction with the molecule neighborhood. Thus,
the transition from the free-space rotationally-symmetric ground state to a constrained situation is an important aspect
in the study of those molecule-neighborhood interactions.
The probability distribution of the molecular orientation in space is one of the keys for the molecule's reactivity with a neighbour
molecule. The knowledge of the molecule's behavior under the influence of charged tips
helps also to design how to switch the molecular axis into some preferred direction.
Though the discussion here focuses on a two-tip-one-molecule system, it might be used in a first step
towards more complicated and extended nanomechanical systems as an example for an incorporated hydrogen molecule. Our example serves to
understand the physical dynamics of the H$_2$ molecule in an external electrostatic field that does not break its bond.

Because of the method's demanding computational costs this idealized configuration still lacks some material specificity at this stage.
Nevertheless, it is related
to realizable sub-nanomechanical physical arrangements and the precision gained here is needed to detect or rule out tiny effects
that might arise beyond the BO approximation or beyond more macroscopic modelling.

\section{Physical System and Computational Method}
\begin{figure}[ht!]
	\centering
	\includegraphics[width=0.8\textwidth,scale=0.5]{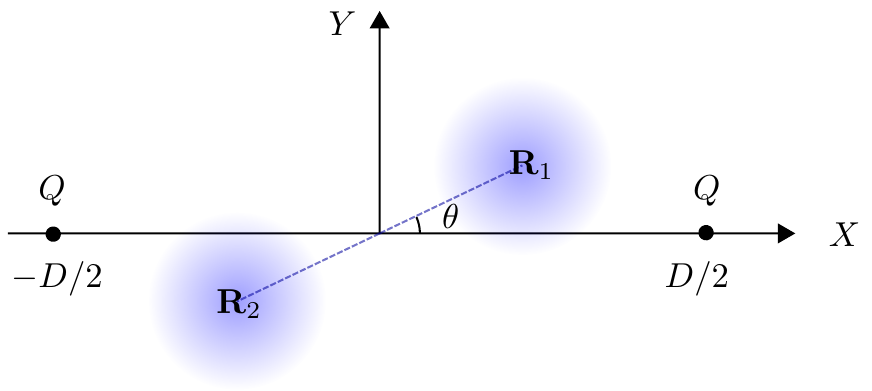}
	\caption{(Color online) Setup and coordinate system for a H$_2$ molecule between
			point charges $Q$ of equal sign, positioned at a distance $D$ apart.
			The molecular axis, defined by nuclear coordinates $\mathbf{R}_1$ and $\mathbf{R}_2$,
			form a polar angle $\theta$ with the $X$-axis.
			We denote the azimuthal angle by $\phi$.
		}
	\label{sketch}
\end{figure}

Our system consists of two tips assumed as point charges $Q$ of equal sign, i.e. $++$ or  $--$, positioned at a distance $D$ apart.
The externally controllable parameters $Q,D$ are varied within the range of a few tenths a.u. for $Q$ and from 4 a.u. to very large distances for $D$. The setup is shown schematically in Fig.~\ref{sketch}. All other
parameters appearing in the calculations are ab initio optimized.
The point charges could have been generalized by replacing them by fictitious H-atoms without gaining much more insight with respect to the goal
of understanding the reaction of the molecule to changing the tip-tip distance and the charge. For simplicity, we consider the center of nuclear masses to be at a high symmetry position, i.e. halfway between the tips. This and the exclusion of the opposite ($+-$) charges is suggestive, as the electrostatic field of the tips on the nuclei already has a saddle point potential shape and the polarized molecule can be expected to attain
an at least meta-stable center of mass position.

With one rotation axis and one mirror plane, this centered system is compatible with two point-like tips of equal charge and has highest symmetry, which usually
is also the fingerprint of a ground state. We checked the stability or instability at that center by means of the energy of the molecule when it is infinitely
distant from the tips. Because of the fixed center of mass the four point charges, electrons and nuclei, need to be quantum mechanically treated
with only 9 degrees of freedom.

This implies that we must here dispense with charge neutrality which would characterize normal environments of a H$_2$ molecule,
such as in a crystal or a liquid. Oppositely charged tips are also of interest because of the voltage drop and the electric current characteristic
in the H$_2$ bridge experiments. It will be pursued in later work.

Nevertheless, the equally charged tips considered here present an interesting variety of physical properties, e.g. the question whether and how
the molecule will dissociate with approaching tips, or whether the molecule will detach as a whole from the tips into the surrounding vacuum.
The field that is produced by the tips in this case also stands for other situations with similar local forces at a hydrogen molecule to which analogously our results regarding stability or vibration frequency can be applied.

Total energy $E$ and Hamiltonian $\hat{H}$ comprise kinetic energy $\hat{T}$, potential energy $\hat{V}$, and interaction energy $\hat{W}$,  observables
of the two nuclei and the two electrons with 6 electron and 6 nucleon degrees of freedom, viz.
\begin{eqnarray}\label{gen1a}
\hat{H} &=& \hat{T}_{nu} + \hat{V}_{nu} + \hat{T}_{el} + \hat{V}_{el} + \hat{W}_{nu-el},\\\label{gen1b}
  &=& \hat{T}_{nu} + \hat{V}^{ad}\\\label{gen1c}
\hat{V}^{ad} &:=& \hat{V}_{nu} + \hat{T}_{el} + \hat{V}_{el} + \hat{W}_{nu-el} = V^{ad}(\hat{{\bf r}}_i,\hat{{\bf R}}_j)\\\label{gen1d}
E &=& \langle \Upsilon|\hat{H}|\Upsilon\rangle = \int \Upsilon^*({\bf r}_i,{\bf R}_j) H \Upsilon({\bf r}_i,{\bf R}_j)d^6r d^6R.
\end{eqnarray}

The general wave function $\Upsilon$ can be decomposed as a product of a part
$\Phi$ depending solely on the nucleons' positions ${\bf R}_j$, $j=1,2$, and a
remaining electron part $\Psi$ which contains the positions ${\bf r}_i$, $i=1,2$, of the electrons,
and additionally,
also depends on the nucleon positions. For example, the centers of the electronic orbitals
represent singularities and their varying position has a non-negligible influence on the nuclear  kinetic energy, in contrast to the simple product ansatz of the BO approximation.
Both parts depend on variational parameters $\alpha_k$
for the electrons and $\zeta_l$ for the nuclei where the former in turn also depend on the nuclei positions.
The nuclear position dependence is represented by the variable distance vector ${\bf R}={\bf R}_2-{\bf R}_1$
between both nuclei, i.e. writing $\alpha_k=\alpha_k({\bf R}_j)$ with implying the presumed fixed center of mass at zero.
\begin{eqnarray}\label{gen2}
\Upsilon({\bf r}_i,{\bf R}_j) = \Psi_{el}({\bf r}_i,{\bf R}_j;\alpha_k)\Phi_{nu}({\bf R}_j;\zeta_l),\hspace*{1cm} i,j = 1,2
\end{eqnarray}
This writing may be denoted as {\it non-adiabatic} in contrast
to the {\it adiabatic} approximation, a notion which refers here to the Born-Oppenheimer (BO) approximation.
Eq.~\ref{gen2} covers the most general eigenfunction of the full electron-nucleon Hamiltonian under a
center-of-mass restriction. Below we sometimes keep the 6 dimensional nuclei space for general writing. We emphasize that
the nuclear positions are quantum mechanical variables such that we do not assign fixed values to them.

The highest symmetry compatible with the geometrical arrangement of two tips and a hydrogen molecule centered in between
is described by cylindrical coordinates. The asymptotic case of the free molecule is reached at infinite tip distance and
yields spherical symmetry. The X-coordinate is taken as cylinder axis and is associated with the longitudinal
motion of the nuclei along the tip's connection line. The ground state nuclear wave function is written in relative
coordinates ${\bf R}:={\bf R}_2-{\bf R}_1=(X,Y,Z)$ as a Gaussian distribution
\begin{eqnarray}\label{gen5}
\Phi^{(0)}_{nu}({\bf R}_1,{\bf R}_2) &=& \exp{[-\zeta_0(\sqrt{(X^2+Y^2+Z^2)}-R_0)^2 -
    \zeta_{0x}(|X|-R_{0x})^2-\zeta_{0yz} (\sqrt{(Y^2+Z^2)}-R_{0yz})^2]}.
\end{eqnarray}
The three parameters $R_0,R_{0x},R_{0yz}$ are used as centers of the
Gaussian ansatz  for the spherical, axial and planar radial dimensions, respectively, instead of a single hydrogen
bond length, and are associated with three corresponding parameters for the widths. They represent
a variational freedom of in total six nuclear parameters which have to be optimized, in addition
to the electronic parameters for the total energy minimum. An ansatz as e.g. the use of the set of Hermite functions instead of Eq.~\ref{gen5} according to
the harmonic oscillator with a ladder of equidistant eigenvalues and one single energy-independent Gaussian exponent is suggestive but too simple. Describing atom-atom interactions one needs to approximate the asymmetry between close and far distances as it is fulfilled e.g. by the Lennard-Jones potential but fails with an harmonic potential, even for the low lying vibrational eigenvalues.
The variational functions obtained here can reflect the actual symmetry on the basis of the nuclear
positions dynamically influenced by the electrons' positions.

Optimizing the electronic variational parameters with respect to a possible dependence on the variable nuclear positions
proves to be necessary and is achieved by sequentially adjusting nuclear and electronic parameters. Besides the above denoted
nuclear parameters, we used 31 electronic parameters that appear in the product of double-zeta Slater-type
orbitals with a Jastrow factor of Gaussian-type orbitals, both formulated within a geminal determinant ansatz~\cite{Casula},
to span the variational space. These 31 parameters more or less sensitively influence the energy expectation
which is used to accelerate the minimum search.

The QMC code, see e.g.~\cite{book_SD}, is organized in the variational mode with runs of $10^9$ electronic moves, every $100$-th of which is
interrupted by a nuclear move. Each run uses a specific parameter realization and is simultaneously repeated on 24 parallel CPUs with
different random number seeds for improved statistics. During optimization, intermediate runs are used with length reduced by a factor of 10 with a respective loss in statistical accuracy by roughly a factor of ~3 for the energy expectation.

\section{Ground State}
\begin{figure}[ht!]
\centering
\includegraphics[width=0.8\textwidth,scale=0.5]{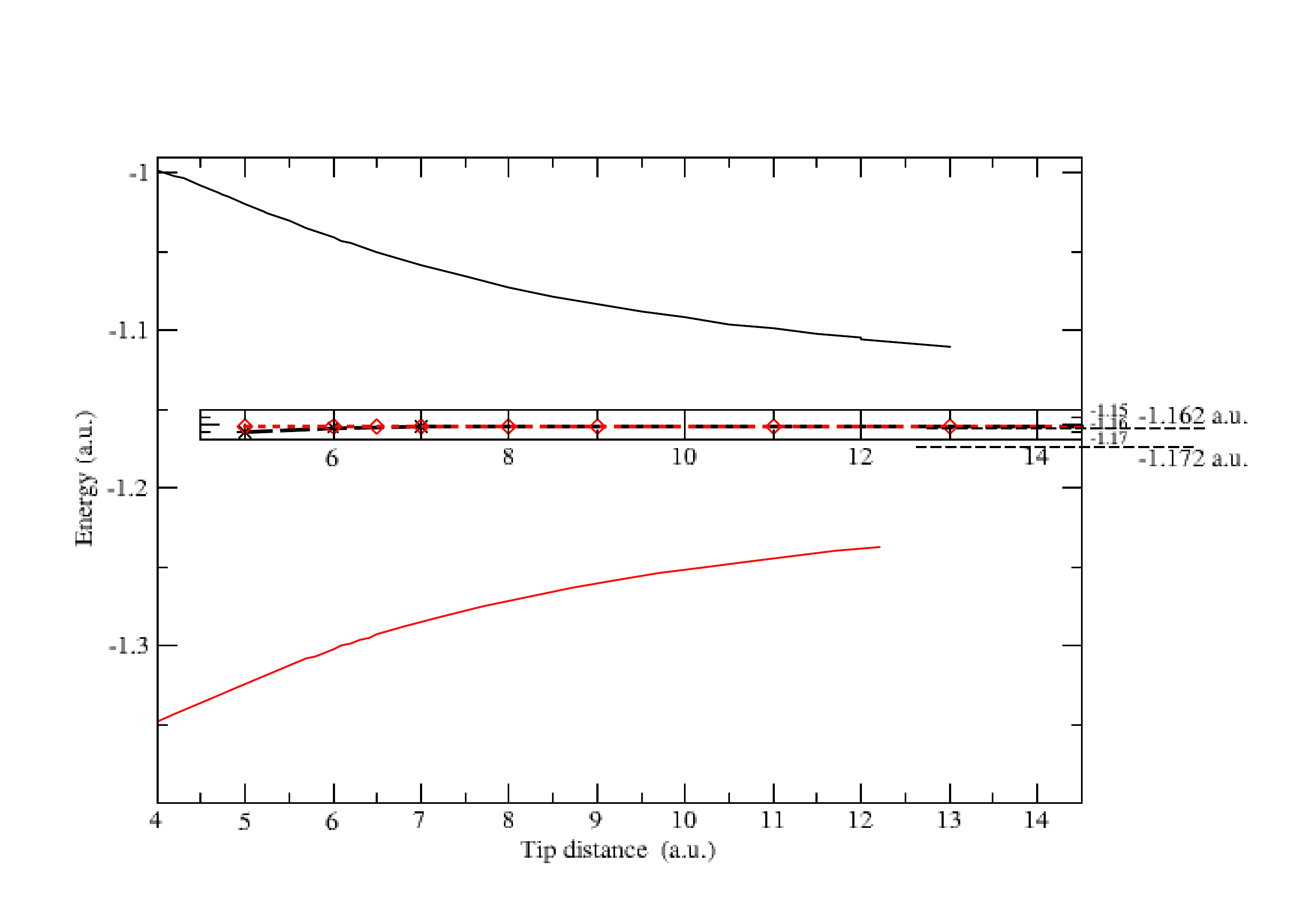}
\caption{(Color online) Sampled adiabatic ground state energy for PES
with varying tip distance and both signs of a fixed amount of equal tip charge Q = -0.1 a.u., lower curve (black), and
Q = +0.1 a.u., upper curve (red); the values obtained here for the free molecule, adiabatic (lower value) and
non-adiabatic (upper value) are indicated at the r.h.s. ordinate with dashed lines,
which might be compared with $E = 1.17448(2)$ a.u. of Tubman et al.~\cite{Tubman}; a preview on the non-adiabatic situation
is depicted in the inset where the axis scales match the outer axes; for explanation see Fig.~\ref{Figurenonadgroundstates}}\label{Figureadgroundstates}
\end{figure}

In the BO approximation we average Eq.~\ref{gen1a} with the truncated wave function of Eq.~\ref{gen2} at static nuclear
positions, i.e. leaving out $\Phi^{(0)}_{nu}$ and optimizing the electronic variational parameters $\alpha_k$ for the potential
energy surface (PES) $V^{PES}({\bf R})$. The latter is the expectation of the observable $\hat{H}$ without $\hat{T}_{nu}$. The ground state
energy is obtained from the quantum mechanical solution of the bare nuclear kinetic energy operator with $V^{PES}({\bf R})$ approximated
by a Lennard-Jones potential. The resulting {\it adiabatic} energy is shown in Fig.~\ref{Figureadgroundstates} as a function of tip distance.
The obtained asymptotic value of the free molecule is shown in addition at the right-hand ordinate to be compared with that of Tubman et al.~\cite{Tubman} where the free case was treated to high accuracy with quantum Monte-Carlo not restricted by the BO assumption. In the adiabatic part of our calculations the molecule was aligned to the tip connection line which reduces significantly the variational space - in the sense of adiabaticity the nuclear positions
are variational parameters - for search of the energy minimum as a
function of the nuclear positions. In fact, as will be seen in the following, the used alignment is totally lost for positive tip charge with
not too distant tips. This adiabatic example presents a caveat and its remedy by the elegant non-adiabatic free floating of the nuclear positions.
It of course does not blame the adiabatic approach itself, since a restricted variational space might include a lower minimum if enlarged.

\begin{figure}[ht!]
\centering
\includegraphics[width=0.8\textwidth,scale=0.5]{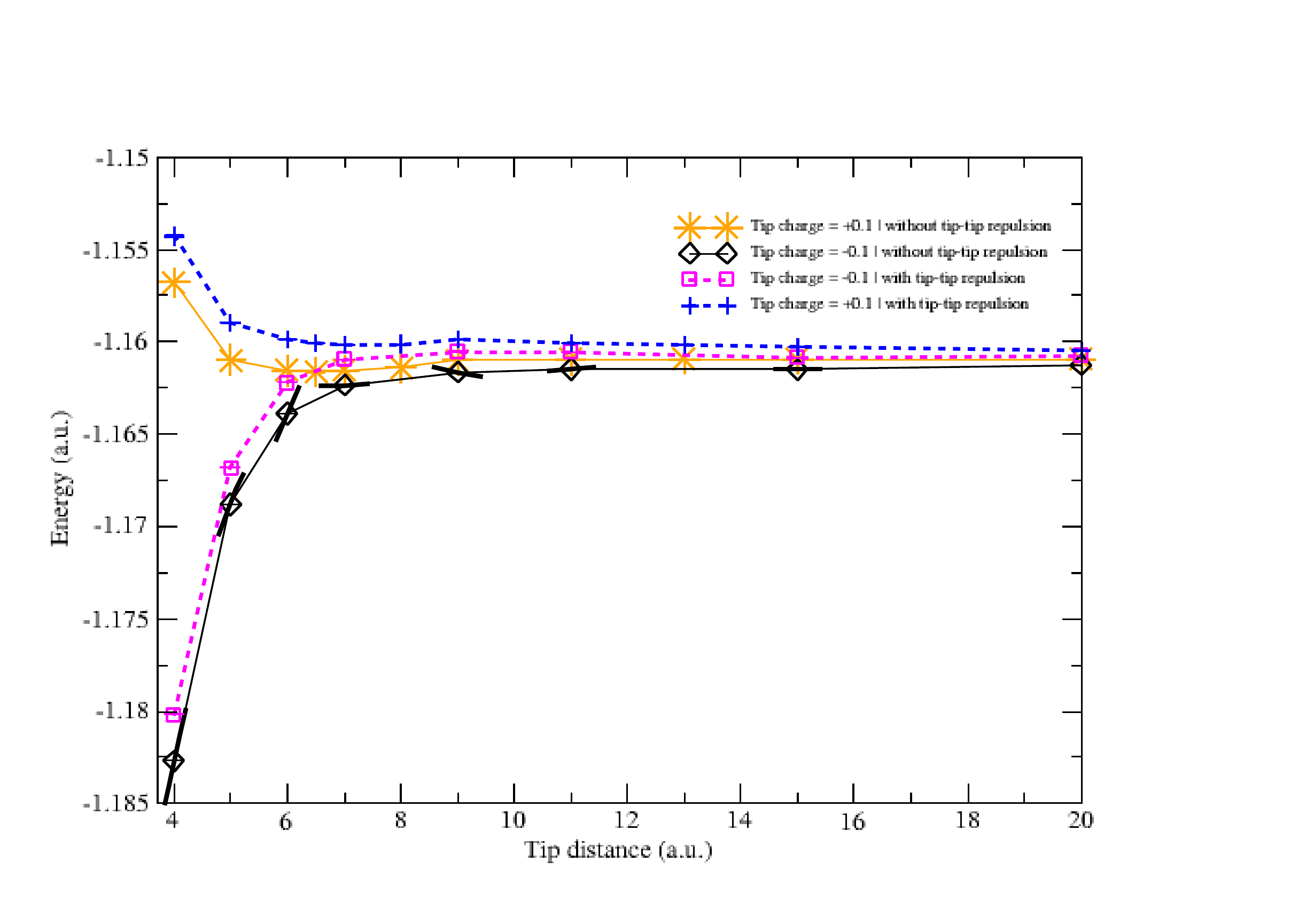}
\caption{(Color online) Non-adiabatic ground states for axial symmetric nuclear wave function ansatz
plotted vs. tip distance for both signs of tip charge: Q = -0.1 a.u. without
tip-tip potential (diamonds,solid) and with tip-tip potential (squares,broken);
Q = +0.1 a.u. without
tip-tip potential (stars,solid) and with tip-tip potential (pluses,broken). Straight line sections at negative
tip charge points represent tangents at those distances obtained from the virial equation,
see e.g.~\cite{book_SD}.}\label{Figurenonadgroundstates}
\end{figure}

\begin{figure}[ht!]
\centering
\includegraphics[width=0.8\textwidth,scale=0.5]{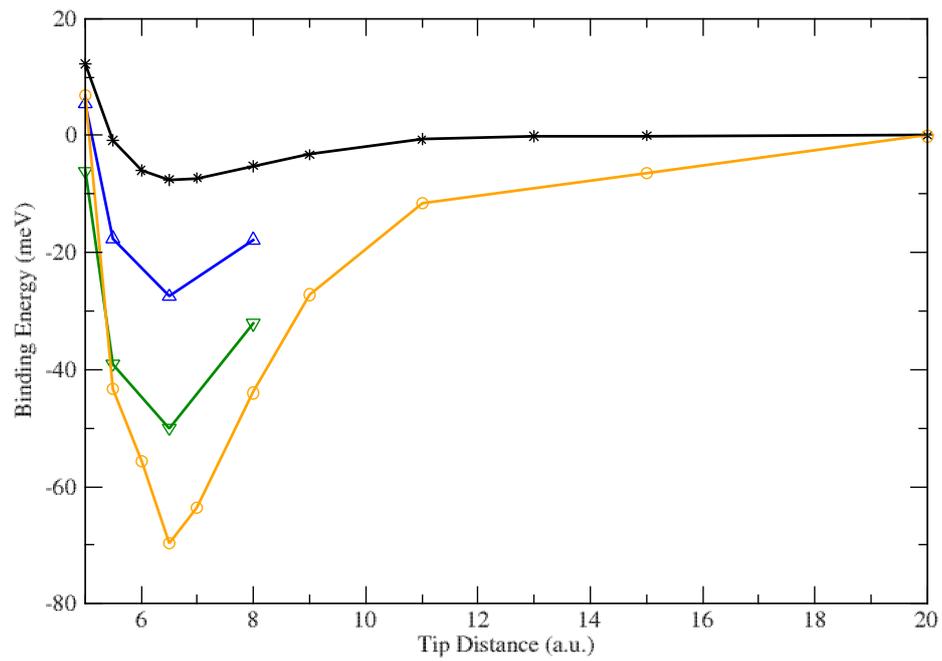}
\caption{(Color online)Ground state energy vs. tip distance for equal positive tip charge of various amounts
$Q=$0.1, 0.2, 0.3, 0.4 a.u. plotted by stars, triangle up, triangle down, and circle, respectively.}\label{AFMvsdistanceandTC+}
\end{figure}

Going beyond BO the full wave function of Eqs.~\ref{gen2} and \ref{gen5} was used for a minimum search of the ground state energy
expectation. The analogue of the adiabatic energy for the {\it non-adiabatic} case is presented in Fig.~\ref{Figurenonadgroundstates}
by solid lines.
Note that the ordinate scales in Fig.~\ref{Figureadgroundstates} and Fig.~\ref{Figurenonadgroundstates} differ by an order of magnitude.
One observes that BO largely overestimates the energy difference between the positive and negative tip charges.
In contrast, the strong stiffness against an external field which the hydrogen covalent bond shows
in the non-adiabatic case can be imagined through the inset in Fig.~\ref{Figureadgroundstates}. Furthermore, a shallow minimum arises
around a tip distance of ~6 a.u. that falls below the large distance level of the almost free molecule. A magnified view is shown
in Fig.~\ref{AFMvsdistanceandTC+} with additional curves for a different value of positive tip charge. With increasing tip charge
the minimum deepens and shows an increase of the binding of the molecule center to the origin. It also proves the stability of the molecule's central position for positive tip charge under the imposed external electric field with respect to its free position far outside.

For negative charge the attractive well becomes significantly deeper, a tendency which could correspond to a separate strengthening of the bonds between each atom and its closest tip neighbor in a kind of physisorption. We calculated the force for checking the accuracy of the plotted data and obtained the tangents at the chosen points as shown in Fig.~\ref{Figurenonadgroundstates}. The derivative vs. tip distance $D$, $\frac{dE}{dD} = -\frac{1}{D}\langle 2T+V \rangle$ follows from homogeneity by a scaling argument as usual.~\cite{book_SD}
By adding the electrostatic tips' repulsion
potential, see dashed lines in Fig.~\ref{Figurenonadgroundstates}, one observes that the complete resulting potential between the tips
proves to be overall repulsive for positive charge, but remains attractive for negative charge. The electrostatic tip-tip interaction is rather smooth
and small compared to the drastic tip-molecule dependence on the charge's sign and size.

Including the electrostatic tip repulsion describes a system governing the dynamical behavior of the tips similar to an atomic force
microscope (AFM) arrangement with H$_2$ between tip and surface. The shallow energy minimum apparent for positive charge and indicating stability
of the molecular configuration within clamped tips vanishes due to the strong electrostatic repulsion if the tips are left free to move,
which then occurs in opposite directions. In the case of negative charge the tip repulsion cannot dominate over the attraction
by the molecule's protons, i.e. cannot balance the intra-molecule attraction over the whole system. The tips will approach, at least
in the region of displayed distances in Fig.~\ref{Figurenonadgroundstates}. At closer approach it can be questioned whether the molecule remains aligned and the bond
eventually opposes further tip approach or turns to a perpendicular direction with decreasing tip-tip distance, thereby decreasing
the proton-tip attraction. This is discussed later in this manuscript. Also,
we recall that the nuclear center of mass is fixed to the origin.

\begin{figure}[ht!]
\centering
\includegraphics[width=\textwidth,scale=0.5]{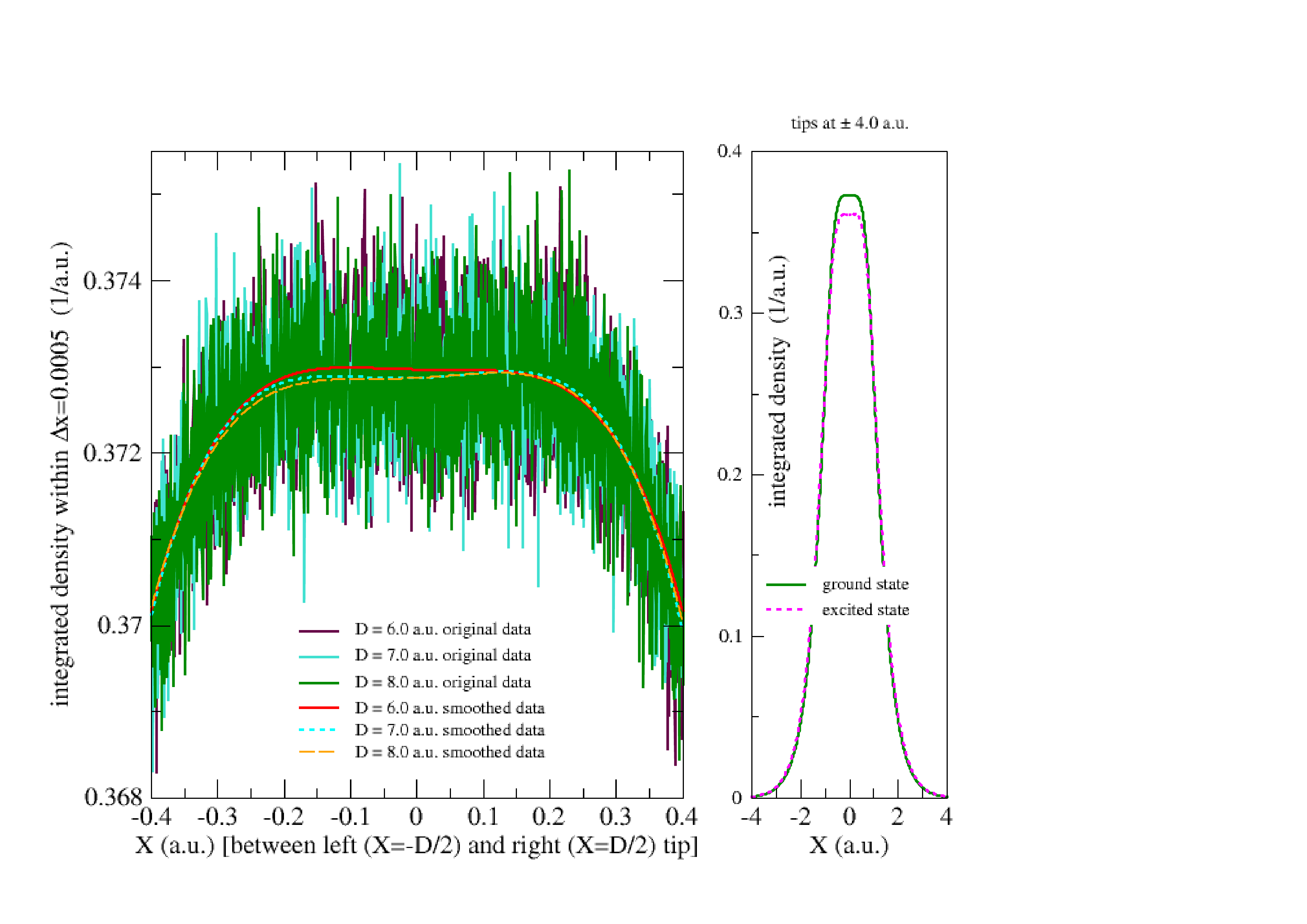}
\caption{(Color online) Electron density (per electron) integrated over $YZ$-sheets
of $\Delta X$ = 0.0005 a.u. thickness. Left: Section around the maximum of original
data plotted by originally scattered and smoothed lines, resp.,
for tip distances of $D$ = 6, 7, and 8 a.u. and tip charge of $Q$ = +0.1
a.u. for ground state of nuclear wave function ansatz of spherical
symmetry, right: Extended curve for $D$ = 8.0 a.u. with both ground
(full) and 1st vibrational excited (dotted) state.}\label{Figureelectrondensity}
\end{figure}

We show
in Fig.~\ref{Figureelectrondensity} a plot of the electronic density for a few distances of the tips. The sampled density is collected
in small sheets perpendicular to the cylinder ($X$) axis. The nuclear wave function ansatz in this example has purely spherically
symmetric parts and a one-dimensional radial nuclear sampling is used, neglecting angular contributions from the nuclear
coordinates' presence in the electronic wave function. Apparently, the electron cloud is arranged over a disk
of average extension equal to the hydrogen bond length along the $x$-axis. Small fluctuations and close resemblance of the curves are observed
in the range of the few tip distances shown, i.e. a rather inert chemical bond at this stage of accuracy. The plateau at maximum electron
density shows two small elevations indicating the H$_2$ nuclear sites' double structure but no visible influence from the varying distance
of the tips.

\begin{figure}[ht!]
\centering
\includegraphics[width=0.8\textwidth,scale=0.5]{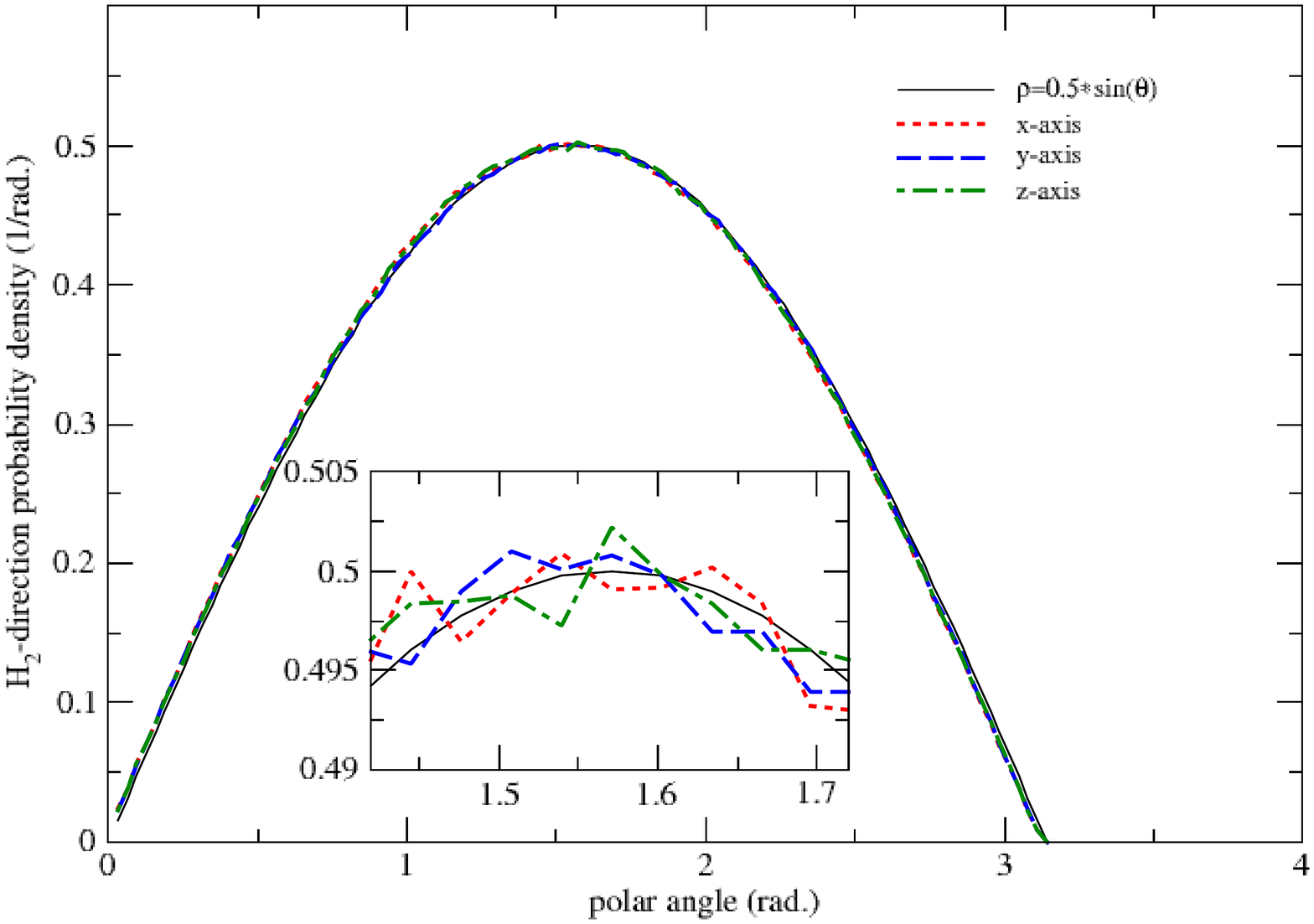}
\caption{(Color online)Isolated (Q=0) H$_2$ molecule: Probability density of polar angles of H$_2$
axis with respect to $X$, $Y$, and $Z$ axes sampled in full three-dimensional space;
amplified resolution shown by the inset.}\label{Figurepolangledist}
\end{figure}

The probability density of the nuclear positions sampled during a Monte-Carlo run illustrates what happens to
the molecular axis for different amounts of tip charge, for central and axial symmetry, and when approaching the tips.
Being a quantum mechanical observable, the orientation of the molecular axis fluctuates. It may be expected to change
if external parameters are changed. The hydrogen axis direction is recorded by the distribution
of its polar angle to each of the three cartesian axes $X$, $Y$, and $Z$. In Fig.~\ref{Figurepolangledist} the probability
density $\rho(\theta)$ for a polar angle $\theta$ is plotted for the case of a spherically symmetric nuclear wave function
of an isolated H$_2$ molecule ($Q=0$) which we still allowed full three-dimensional degrees of freedom in the sampling.
The directional fluctuations from a homogeneous distribution of $\frac{1}{2} sin(\theta)$ are small, as is illustrated by a magnified plot
section in the inset.

\begin{figure}[ht!]
\centering
\includegraphics[width=0.8\textwidth,scale=0.5]{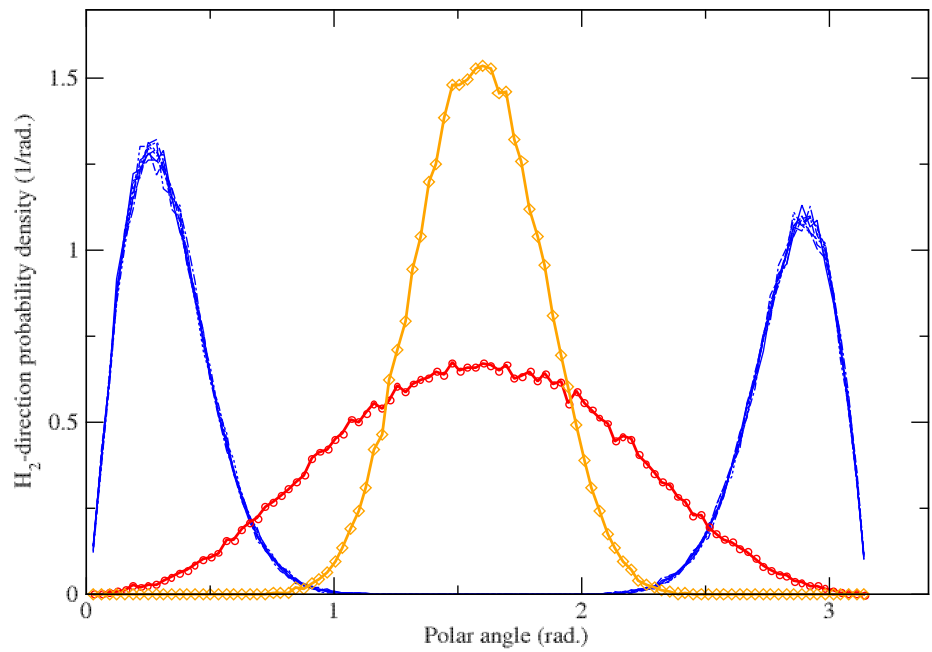}
\caption{(Color online)H$_2$ molecule embedded between tips (Q $\neq$ 0): Equatorial and conical probability distributions of molecular axis vs. polar angle in the strong axial symmetry regime of the ground state:
tip charge $Q$ = -0.1 a.u. with thin solid and dashed lines (blue) for a set of azimuths $\phi$ = 0.031, 0.063, 0.094, 0.126, 1.508,
1.539, 1.571, and 1.602 rad. for tip distance $D$ = 5.0 a.u.; tip charge $Q$ = +0.1 a.u. for azimuth $\phi$ = 0.031 rad.
with solid line (orange, open diamond) for $D$ = 5.0 a.u. and solid line (red, open circles) for $D$ = 6.0 a.u.}\label{Figurelateralconicalangle}
\end{figure}

In contrast, with finite tip charge a drastic change is observed when the positively charged tips are approached: A symmetry breaking increasingly takes place
towards axial symmetry. The electrostatic repulsion between tips and partly unscreened nuclei turns the molecular axis
distribution towards the equatorial plane. The turning of the H$_2$ axis follows qualitatively this electrostatic picture according to the tips'
field distribution whereas this illustrative explanation quantitatively needs a strong electronic screening of the proton charge to reduce the
electrostatic energy by an order of magnitude to the QMC computed quantum mechanical value.
The emerging turning of the molecular axis is illustrated by Fig.~\ref{Figurelateralconicalangle} where both cases of tip charge sign are displayed.
The presentation is slightly different from that in Fig.~\ref{Figurepolangledist} with the reference polar axis fixed along the tips connection line ($X$ axis), the main symmetry axis. The abscissa here denotes the polar angle $\theta$ with respect to the $X$ axis, - and a few values of the azimuth angle $\phi$
appear as plot parameter -. The $\theta$ accumulation around $\pi/2$
for positive tip charge indicates a preferred direction near the equatorial plane. The two-lobe distribution then corresponds to the two
opposite cones into which the internuclear distance vector can point in the case of negative tip charge. The azimuth distribution is apparently uniform,
as it should be.
The slight asymmetry towards a more frequent pointing to the positive hemisphere originates from the initial start position
of the sampling. Note the implicit sine-factor weight from the azimuthal degree of freedom present
in the distribution that forces the distribution to zero at both borders.

Aside from the very pronounced depression in the case of negative tip charge, where a large attraction after an initial repulsion is seen for decreasing tip distance (see Fig.~\ref{Figurenonadgroundstates}), a barely noticeable small dip appears in the ground state energy of positive tip charge between large and small tip separation. The amount of the depression, however, increases for larger tip charge and could be resolved by an AFM.
We investigated larger positive tip charges, see Fig.~\ref{AFMvsdistanceandTC+}, showing that the shape of curves together with the position of their depression minima are rather independent of charge. The depth appears to increase rather linearly with tip charge such that energies in the range of a few hundred $meV$ are attainable. We thus observe already in this simple system a trapping of the molecule that has been stated by Hui Wang et al.~\cite{HuiWang} for nanocavities built by an atomic environment at a surface. Additionally, the trapped H$_2$ in this potential well, where
the tip distance can be mapped onto the tip bias, experiences a bias dependent varying force and may by this nonlinearity couple to an external oscillator connected with the tip, as realized in an elegant design~\cite{Lotze} for a H$_2$-covered Cu(111) surface.

In other words, for both signs of charge there exist regions of total energy below its value for the free molecule, thus proving stability of the center of mass position at the origin.
\section{Vibrational Excitation}
In this contribution we consider the molecular vibrations as key to measurement. Variational QMC schemes are primarily devoted to ground state
investigations~\cite{book_SD}; however, the conceptual simplicity of the nuclear wave function reduces the variational efforts for excited states
by providing near orthogonality through symmetry arguments. As the ground state has highest symmetry, we choose, similar to the harmonic case, an ansatz antisymmetric in the stretching mode coordinate for the first vibration amplitude. This ansatz is motivated by assuming a smooth transition
from the free molecule with spherical symmetry to its constrained symmetry with one remaining rotation axis. And, the choice of a spheroid suggests itself from the ground and excited state exponent's elliptic expression as variational ansatz. In this way, cylindrical symmetry of the whole system is obeyed; however, it complicates the notion of vibrational modes of free binary molecules.
In fact, if we refer to the bridging experiments and focus on the stretching mode,
a purely radial eigenmode
does not exist in cylindrical symmetry. Instead, aside from the rotational eigenmodes around the tip-tip axis, the other modes generally contain a mixture of a stretching vibration and a rotation with polar angle referring to the tip-tip axis. Besides,
the term stretching mode now corresponds in the quantum mechanical continuous space description to a breathing mode in purely spherical symmetry with additional distortions by differential rotation in axial symmetry.

The surface of the wave function zero is adapted from the free molecule's spherical shape to
a spheroid as function of the 3-dimensional nuclear relative coordinate ${\bf R}={\bf R_2}-{\bf R_1}$.
This assumption complies with the axial symmetry and a perpendicular mirror.
It is further specified by adjusting a simple quadratic polynomial to the expected linear behavior
in the neighborhood of the zero at the spheroid. Two cases are encountered according to the ground state findings: the prolate spheroid elongated along the tip-tip direction, as e.g. for negative tip charge and the oblate spheroid flattened perpendicular to that direction for positive charge. From now on, we focus on the former case, as it possesses a large stability region, and write
\begin{eqnarray}\label{gen8}
\Phi^{(1)}_{nu}({\bf R}_1,{\bf R}_2) &=& (X^2+\frac{Y^2+Z^2}{1-e^2}-a^2)
    \exp{(-\zeta^{(1)}_x(X-R^{(1)}_x)^2-\zeta^{(1)}_\rho(\sqrt{Y^2+Z^2}-R^{(1)}_\rho)^2-\zeta^{(1)}(|{\bf R}|-R^{(1)})^2)}
\end{eqnarray}
as the nuclear factor of the vibrational excited wave function. The spheroid is characterized by its semi-major axis $a$ and
numerical eccentricity $e$ as variational parameters together with $R^{(1)}, R^{(1)}_x, R^{(1)}_\rho$ for the variational centers
and $\zeta^{(1)}, \zeta^{(1)}_x, \zeta^{(1)}_\rho$ for the variational widths of the Gaussian ansatz. The oblate case would have the
$\frac{1}{1-e^2}$ factor at the $X^2$ term instead.
Orthogonality to the ground state is imposed by integrating the mutual projection $(\Phi^{(1)}_{nu},\Phi^{(0)}_{nu})$
of the respective nuclear factors and choosing the semi-major axis value to set this scalar product to zero.

The eccentricity can be treated as a parameter either free for energy-variance minimization or prescribed by assuming the vibration
polarization in advance by some special cases. The latter are considered here by a sphere with $e=0$ for a quantum mechanical breathing mode corresponding to the classical stretching mode, a plane with $e=1$ and node $Y=0$ for a ro-vibration mode of pendulum type in the $(X,Y)$-plane, and a twofold plane $|X|=X^{(1)}$ of also pendulum type but with azimuth symmetry around the tips' axis. As a first guess, the constant $X^{(1)}$ is chosen
to fit the center of the quadratic expression in the nuclear wave function exponent different from orthogonality requirement as described above. By the specific choices of the nodal surface we
introduce a confinement of the nuclear Hilbert space which restricts its low lying ro-vibrational excitations. We expect them
playing a role in the inelastic electron transfer of the hydrogen bridge experiment as transition states weighed by respective matrix elements.

In principle,
the integration over nuclear coordinates should also comprise its respective dependence in the electron wave function part e.g. by
a Monte-Carlo sampling in a separate step. Its influence on the nodal surface is neglected here in view of other remaining uncertainties
as e.g. slight shape corrugations of the unknown exact nodal surface.

\begin{figure}[ht!]
\centering
\includegraphics[width=0.8\textwidth,scale=1.0]{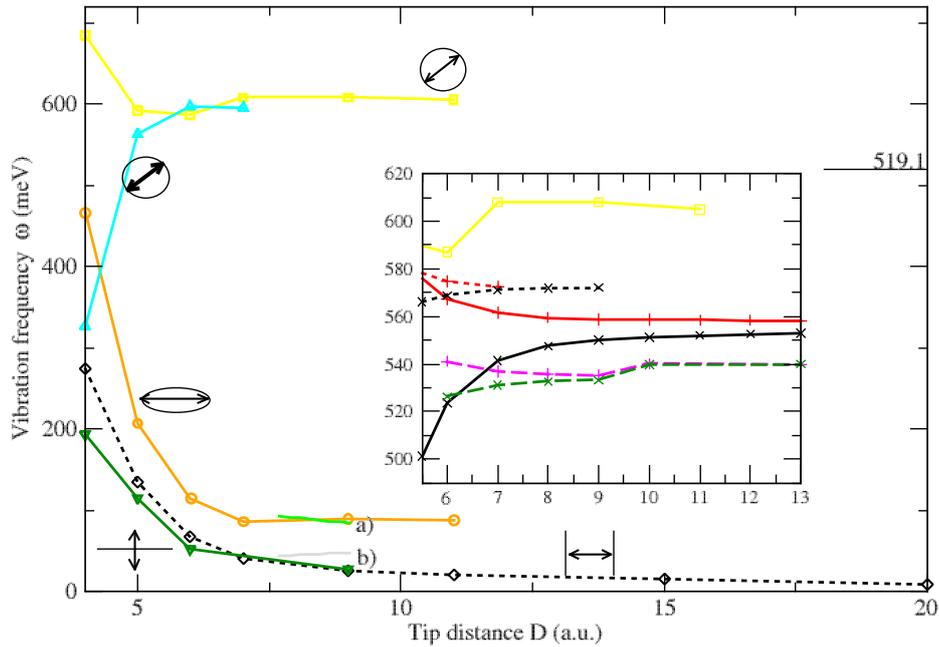}
\caption{(Color online) Non-adiabatic lowest vibration excitation  frequencies vs. tip distance for various node surfaces: pure  stretching mode (squares,solid,yellow) with spherical nodal surface, pendulum mode (circles, solid, orange) with optimized spheroidal nodal surface, pendulum mode (triangles down, solid, green)
with node on plane $y=0$, and pendulum mode (diamonds, dashed, black) with node on plane $|X|=constant.$ all for tip charge $Q=-0.1$ a.u., and the pure stretching mode (triangles up,solid,cyan) for tip charge $Q=-0.2$ a.u. with spherical nodal surface.
Inset shows a section of the stretching frequencies plotting the spheroidal ansatz for a 3-dimensional spherical node, (squares,solid,yellow) as above, in comparison with  1-dimensional calculations of the radial nuclear distance coordinate assuming full spherical symmetry: on the basis of BO with PES and a fitted Morse potential
after QMC electronic parameter optimization
for tip charge $Q = +0.1$ a.u. (pluses, solid, red) and $Q = -0.1$ a.u. (crosses,solid, black), on the basis on the harmonic approximation
(dotted, red/black) for both charges $\pm$ 0.1 a.u. resp., and on the basis of the  non-adiabatic treatment in that enhanced symmetry $Q = +0.1$ a.u. (pluses, dashed, magenta)
and $Q = -0.1$ a.u. (crosses, dashed, green). Literature value~\cite{Dickenson} is marked at right ordinate. Icons with arrows near a curve symbolize vibration polarization mode by its  shape of the nodal surface projection onto $(X,Y)$-plane as circles and ellipses, the latter also with its degenerate linear shapes $Y=0$ and $X=\pm{X^{(1)}}$. Experimental curves~\cite{Djukic} of figure 3 therein, right shifted by 9 a.u., are depicted by two almost straight lines with marks a) and b).  }\label{Figureomegavstipdistance_general}
\end{figure}

The local velocity at and close to this nodal surface is directed normal to the surface along the intersecting confocal hyperbolas
within any plane containing the tip-tip axis. It is composed of a radial stretching vibration and a rotational component which increases
with increasing eccentricity $e > 0$. The additional rotation is due to the tips' symmetry breaking
with an energy amount corresponding to the tip charge, roughly speaking. The degree of freedom connected with the polar
angle mixes thus into the stretching vibration eigenmode. As the low rotational mode frequencies are smaller by an order of magnitude, one could
expect the mixing only slightly decreasing the vibration frequency for small perturbations by the tips' field.
Altogether the field becomes strong for very small tip distance, a regime where the hydrogen bond already starts to break.
In that regime, the field appreciably pulls down the vibration frequency anyway as seen in  Fig.~\ref{Figureomegavstipdistance_general}, and arising as a general feature
when compared with various assumptions about wave function approximations in the inset.

\begin{figure}[ht!]
\centering
\includegraphics[width=0.8\textwidth,scale=0.5]{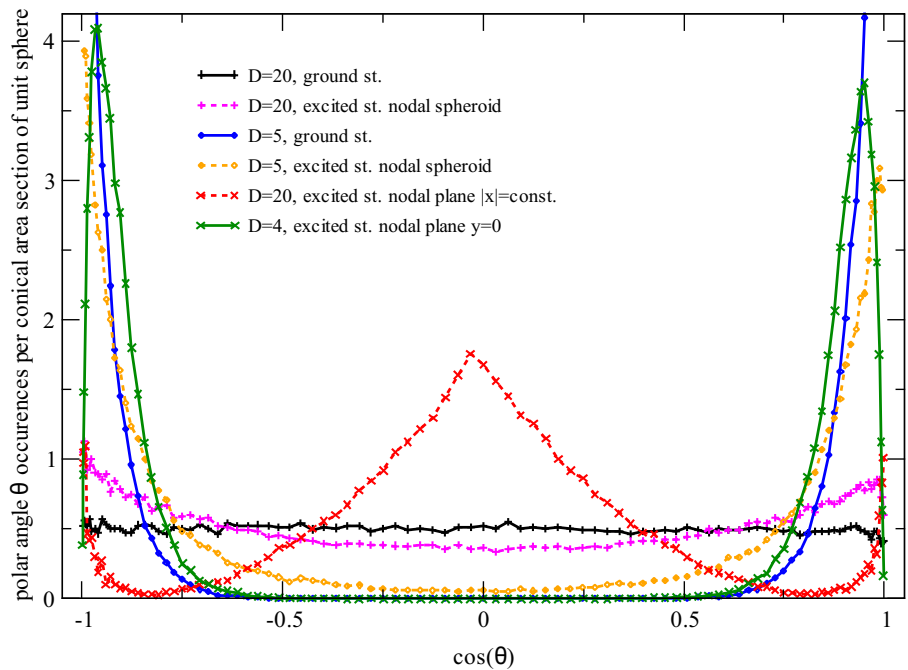}
\caption{Normalized occurrences of polar angle between tips' connection and hydrogen axis vs. cosine of polar angle $\theta$. Occurrences refer
 to the ring area on the unit sphere between $\theta$ and $\theta+d\theta$: ground state (solid) for tip distance $D =$ 20 (pluses,black) and for $D =$ 5 a.u. (diamonds,blue), vibrational excited state for stretching mode on a spheroidal nodal surface (dashed) with $D =$ 20 (pluses, magenta) and 5 a.u. (diamonds, orange), and rotation-vibration excited state with nodal plane at
 $Y=0$ for one transversal mode (solid, crosses, green) and at $|X|=const.$ (dashed, crosses, red).}\label{Figureconical occurrences}
\end{figure}

Considering the spheroidal nodal surface in the case of negative tip charge, we expect from the ground state wave function shape
a prolate spheroid along the tips' connection line. It is interesting to investigate for the nearly spherical symmetry at tip distance of 20 a.u. also the extreme opposite choice mentioned above for excited state: the
oblate spheroid with eccentricity $e=1$, which is a twofold planar nodal surface at $\pm X^{(1)}$. It is obtained by replacing the spheroid prefactor of the exponential
in Eq.~\ref{gen8} by $(|X|-X^{(1)})$
and choosing $X^{(1)}$ by reference to the harmonic approximation disregarding a possible  non-orthogonality to the ground state. The resulting distance behavior of the vibration energy is plotted in
Fig.~\ref{Figureomegavstipdistance_general} and shows for this case an almost vanishing frequency at large distance. The directional
distribution of this special case as displayed by the plotted occurrences in Fig.~\ref{Figureconical occurrences} for $D=20$ a.u.~and $Q=-0.1$ a.u. shows a large accumulation in the equatorial plane and peaks at both poles with a node-like behavior in between. The
latter lies roughly in the region, where it appears by the formal node ansatz of the excited state, namely as polar angle at the intersection between the sphere of maximum exponent and the nodal plane $|X|=X^{(1)}$.
One can consider those two maxima at pole and equator in the angular probability density as the wave function´s fingerprint of the lowest vibration excitation. Speaking in simple classical terms, one may see a large rotation of ~90$^{\circ}$ being involved in the vibration, when the nuclear distance vector slips over this angle from its minimum to its maximum length. The corresponding long nuclear path
forces the frequency to a very small value. Thus, the smallest quantized angular momentum with respect to the $|X|$-axis polar angle can be asymptotically ascribed in the corresponding spherical limit. This $\Pi_X$ type of state shows rotation symmetry around the $|X|$-axis which adapts it to an inelastic transition from the ground state at least by its equal symmetry. However, it does not reflect or even approximate a stretching vibration which should be radial, of course, and searched in a spherical approximation to the nodal surface.

The variational states here obtained are eigenstates, i.e. long living, within our accepted range of accuracy given by a square root value around 0.05 a.u. of local energy variance and around 0.000005 a.u. of total energy expectation variance. Calculating the inelastic electron transfer observed in
experiment must determine an acceptable excited state by the transition matrix element aside from its ground state orthogonality. As a rough estimate we consider in addition to symmetry selection rules the directional overlap of the excited state with the ground state including an additional one-electron orbital on both tips being occupied, say $\phi_R$ on the right-hand side in the ground state $\Phi^{(0)}$ and $\phi_L$ on the left-hand side in the excited state $\Phi^{(1)}$. For example, a transfer matrix element $(\phi_L\Phi^{(1)},V_D\phi_R\Phi^{(0)})$ has to be investigated for a hydrogen-assisted or hydrogen-blocked inelastic tunnelling using the tunnelling theory~\cite{Bardeen}, STM~\cite{Tersoff}, or a scattering $t$-matrix~\cite{Birkner} for impurity assisted tunnelling processes. The driving potential for an electron transfer $V_D$ consists of a voltage drop between the tips illustrated e.g. by an additional static negative point charge on the right tip. It has also rotation symmetry with respect to the $X$-axis, thus admitting e.g. the $\Pi_X$ type state described above at least by symmetry.

Examples
for two tip distances are displayed in Fig.~\ref{Figureconical occurrences} showing the similarity in the axes' angle distribution
for ground and excited state. Such a similarity increases the transition probability from ground to excited state and would favor
an inelastic electron transfer via a hydrogen bridge, provided that the operator in the matrix element depends only smoothly on the angle.
Conversely, the $|X|=const.$ planar nodal surface has less overlap with the ground state at a tip distance of 20 a.u.,
and, additionally, shows this angular overlap around the equatorial plane which the orbital of a tip electron rarely could access
in order to be able to tunnel.
These qualitative arguments do not strictly exclude other parts from the large excited states' space of the rotation-vibration spectrum
even at the low excitation level of some of hundreds meV. Even the excited state with planar nodal surface, which has been ruled out above, shows similarity with the bridge experiment insofar that its energy is monotonous decreasing with increasing tip distance.
The leading selection considered here is based on the orthogonality to the ground state and the continuity with the isolated molecule.
A definite answer would need a calculation of the transition matrix element and is beyond the scope of this work.

Before discussing the final vibration frequencies of our calculations we show in the inset of Fig.~\ref{Figureomegavstipdistance_general} the frequency dependence on tip distance for three approximate schemes,
namely a harmonic approximation to the sampled BO PES, a Morse potential fit to the PES itself,
and a non-adiabatic calculation with the assumption of spherical symmetry of the nuclear wave function.
The vibration frequency softens (hardens) for negative (positive) tip charge with increasing external field strength
by approaching tips. All those results based on spherical symmetry are not influenced by the turning molecule axis
that has been found in the ground state calculations as a rotation into the equatorial plane for positive charge and as an aligning with the tips for negative charge.
Such differences in configuration are expected to affect the vibrational modes, and the process of their excitation.

In fact, the actual axial symmetry forces a hybridization of vibration and rotation of an eigenstate. For example, the transition from the free molecule with radially symmetric nuclear wave function to the molecule bond by the tips in cylindrical symmetry gradually restricts the molecule's motion within some cone around the tip connection line or within an equatorial disk depending on charge.

\begin{figure}[ht!]
\centering
\includegraphics[width=0.8\textwidth,scale=1.0]{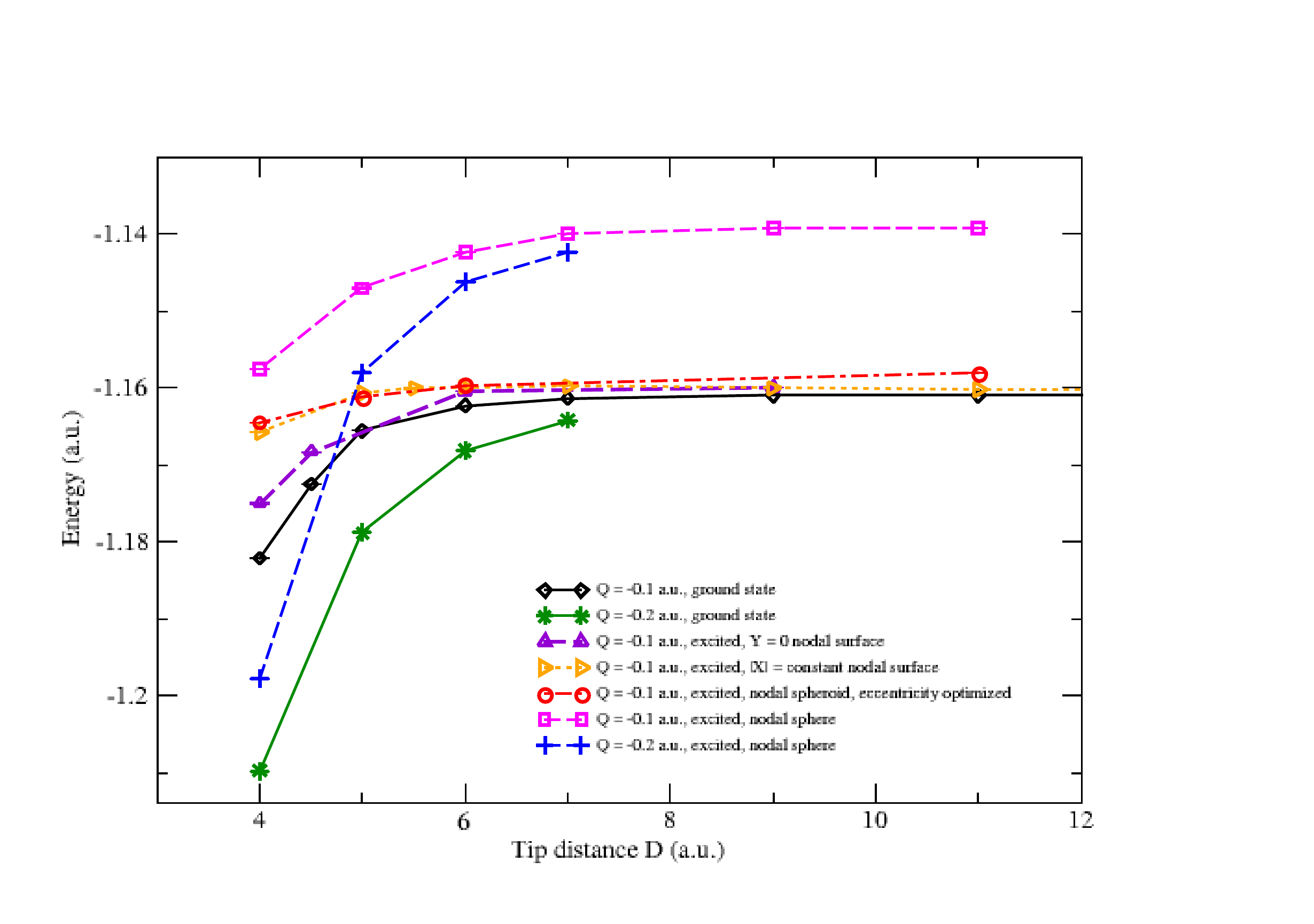}
\caption{(color online) Energy vs. tip distance: ground state curves for $Q = -0.1$ a.u.(diamonds, solid, black) and $Q = -0.2$ a.u.(stars, solid, green); vibrational excitation curves (broken) for several nodal surface shapes: spherical surface for $Q = -0.1$ a.u.(squares,magenta) and for $Q = -0.2$ a.u.(pluses, blue), and three curves for $Q = -0.1$ a.u. the general spheroidal shape  (circles,red), a two plane set perpendicular to tip-tip connection at $X = \pm X^{(1)}$ (triangles right, orange), and the $(X,Z)$-plane $Y = 0$ (triangles up, violet).}
\label{Figurenonadexstenergy_vs_TD}
\end{figure}

As a stable trapping of the molecule for positive tip charge is restricted to a rather small tip distance range, we consider as mentioned above more closely only the negative charge case
with the enhanced parameter space of the spheroid nodal surface. In Fig.~\ref{Figurenonadexstenergy_vs_TD} the excited states' energy vs. tip distance is plotted together with
the ground state curves. There are 4 excited states' curves plotted as low lying final states
above the $Q = -0.1$ a.u. ground state:
\begin{enumerate}
    \item The uppermost curve corresponds to a sphere as nodal surface, the classical stretching
    mode.
    \item In optimizing the eccentricity $e$ of the general spheroid nodal surface the state appears at lower energy with a dependence similar to and asymptotically approaching that of the next item below.
    \item The state with two nodal planes at the poles $\pm X^{(1)}$ corresponds to the configuration already described above. The angular dependence is symmetric around the $X$ axis and is referred for far distant tips as $\Pi_X$ nuclear state because
    of this asymptotic angular symmetry.
    \item Below and close to the $\Pi_X$ curve an eigenstate of the $Z$-component of angular momentum is plotted with the nodal plane at $Y = 0$ and denoted as $\Pi_Z$ state which corresponds to a classical pendulum mode where the hydrogen axis swings in the $(X,Y)$ plane.
\end{enumerate}
The last mode which has, of course, a degenerate counter part ($\Pi_Y$, not shown) in the $(X,Z)$ plane shows the axis distribution as visualized for tip distance $D = 4$ a.u. in Fig.~\ref{Figureconical occurrences} entirely focused to the tip connection with almost vanishing lateral probability. This behavior is in clear contrast to the $\Pi_X$ mode which shares its main probability between the equatorial plane and the poles with a
separating angular node. The nodal spheroid, 2nd item, shows in both cases $D = 5$ and $D = 20$ appreciable wave function directional overlap of ground and excited state to allow for a coupling to an electron transfer through a hydrogen bridge. This curve is optimized and therefore more reliable,
though it lies to a small extent above the $\Pi_X$ state in Fig.~\ref{Figurenonadexstenergy_vs_TD} as the latter lacks strict orthogonality to the ground state. Both the $\Pi_X$ and the $\Pi_Z$ state can be viewed as the lowest excitation according to an angular momentum expansion associated with the $X$ and $Z$-axis resp. as polar axis that could be expanded through higher momenta towards higher accuracy.

The case of a spherical node becomes special for very close tips as the nodal surface then is extremely different from the constant energy surfaces of
the exponential part of the nuclear wave function and more important, its energy lies far above the lowest excitation obtained from general spheroidal optimization. In view of the vibration energy we calculated  this case also for doubled charge $Q = -0.2$ a.u.. The excited energy displays a similar shape as that of the ground state in all cases. Remarkably, it falls below the non-adiabatic energy
level of the free molecule ground state except the $Q = -0.1$ a.u. nodal sphere.

The difference between excited and ground state yields the vibration energy plotted in
Fig.~\ref{Figureomegavstipdistance_general}. The spheroidal eigenmode is shown to soften appreciably with decreasing tip distance $D$. It approaches with increasing $D$ both the $\Pi_Z$ pendulum mode and the limiting $\Pi_X$ rotation-vibration mode which themselves approach each other still closer. The spherical mode shows a slight bending downwards with decreasing $D$ from large values for $Q = -0.1$ a.u. and a much stronger bending for $Q = -0.2$ a.u. with an onset already at higher $D$. This behavior resembles the shape of the stretching curve of figure 4 of Djukic et al.~\cite{Djukic} in its upper portion with $D > 4$ a.u. if $Q$ is adjusted here to somewhat larger negative charges. The curves we find for the pendulum and the rotation-vibration modes in Fig.~\ref{Figureomegavstipdistance_general} have a shape which could with some imagination be associated  there in figure 4 with the tail of the transverse pendulum mode excluding
the center of mass modes. Here, we identify them by a purely lateral motion along the $Y$  or longitudinal motion along the $X$ axis with fixed center of mass. Thus, the association of our curves with density functional theory (DFT) calculations and experiment bears certain ambiguity because of the notion of polarization in terms of classical or quantum mechanics, on one hand, and because of the different tip distance ranges used, on the other hand.

The QMC calculations may be also directly compared with the experiment though the measurement does not specify a polarization signature  of the wave function with respect to the vibration eigenstate. One can try to associate in that case the shape of curves with specific excited states from QMC, namely a spherical breathing mode and a spheroidal mode where breathing is mixed with shearing. Because of its high excitation energy the breathing mode we obtain here is not accessed by experiment. Our curve with doubled
tip charge $Q=-0.2$ a.u. tends to lower the energy which, however, occurs at too small a tip distance. We don’t know what the zero of the experimental abscissa is, so we feel free to adjust it yielding the curves denoted a) and b) in Fig.~\ref{Figureomegavstipdistance_general}. This shall not suggest an approximation to experiment as our external parameters $(D,Q)$ would have to be realistically mapped onto the measured  situation. But it gives an impression which portion of tip distance might correspond to the rather small measured region. The value of the frequency as well as the slopes of the curves suggest that tip distances above 7 a.u. can probably be associated to our spheroidal mode with one of the curves a) or b).
The cited DFT calculations - being supported by our own calculations on a linear chain of Au atoms with a gap bridged by a hydrogen molecule - yield in addition the wave function's finger print as being longitudinal or transverse and with center of mass being fixed or oscillating.
Our DFT results indicate that the H$_2$ bond breaks below 10 a.u. which compares with 7 a.u. of QMC
where the hydrogen bond starts to significantly weaken. At this distance, the main $s$ orbital appreciably broadens with respective bond length increase.
Thus, there is correspondence between DFT stretching and QMC breathing modes. The remaining pendulum mode is rather unspecific and not easily identified. At least it does not contradict the pendulum mode of DFT. The difference could originate from the use of the BO approximation in DFT as opposed to
the QMC non-adiabatic treatment with a basically different concept of correlation. But it could as well be a simpler reason buried in the correspondence between both models, i.e. between the external parameters $D,Q$ of the QMC bridge model with point-like tip charge and tip distance and the DFT model with atomically structured tips.

The findings in the bridge~\cite{Djukic} and cavity experiments~\cite{Lotze},\cite{Carbonell} for the vibration
energy vs. tip distance curves show some similarity to the data presented here. Two  curves in figure 3 of Djukic et al.~\cite{Djukic} deserve special consideration, as they are in the reach of the parameter space covered here.
\section{Conclusions}
For the geometry of the system which we have studied here, QMC calculations have been performed to obtain the total ground state energy for
both signs of charge. The results show overall stability for a negative tip charge. A charge of equal but positive amount
yields stability with respect to the free molecule for tip distances larger than around
$D = 5$ a.u. The deviation from the BO evaluation
can be estimated by the energy difference of both charge signs between the PES and the non-adiabatic case that
amounts up to 0.3 a.u. at about $D = 10$ a.u. The quantum mechanical concept of the protons' wave function significantly
reduces the differences appearing
in a classical treatment within the BO frame. In addition, the molecular orientation adjusts itself to an external field yielding
besides the strong reduction of the energetic difference also an unexpected qualitative difference in the proton-proton angular wave function
distribution. The probability distribution of molecular orientation in space is important
for the molecule's reactivity with a neighbouring molecule. It can be influenced by such a tip configuration and switched by the tips' charges
into an a priori preferred direction. A technical by-product of the present work proved
to be the necessity of some repetitions in optimizing in turn the variational parameters of the nucleon-only part and those of the
electronic wave function part which is beyond a variational BO decoupling. Total energies could already be improved by around 100 meV by the second optimization of the electronic parameters after the
nuclear parameters.

Though the discussion focussed on a two-tip-one-molecule system, it might be used in a next step towards more complicated and extended
nanomechanical objects as an example for an incorporated hydrogen molecule with electrically induced switching properties.

The molecule can be examined by detecting its vibrational spectrum. The results on the first vibrational excitation show that
the energy can be resolved by QMC. Qualitatively a similarity emerges between the full quantum mechanical and a purely electrostatic view
of the proton dynamics for suitably screened charges.
The quantitative results for the free molecule are determined with the high statistical accuracy (usually line widths in the plots)
of the QMC scheme. The systematic error can be asymptotically controlled by reference to the free molecule value. The absolute value
of the ground state energy is estimated to lie about 60 meV ($\sim 10$ \%) above the exact value for the free case, whereas the vibration
frequency is found to deviate by about 70 meV ($\sim \pm 10$ \%) relative to the values of the free molecule.

A variety of curves of similar shape are shown
in the graphs of vibration frequency versus tip-tip distance for various approximations. In addition, assumptions are necessary
about the actual vibration modes coupled and being probed in an experiment with external electrostatic constraints,
be it the free spherically symmetric stretching mode or the axial vibration in cylinder symmetry which becomes soft by the hybridization of stretching with rotation. For example, the frequency vs. distance shape will highly depend on the coupling matrix elements exciting
the tunnelling electron in the bridge experiments.

In any experimental realization, equal positive charges on both tips designed as above works similarly to an anode, whereas negative test charges
could be offered to the molecule by one or more additional tips in the mirror plane as cathode. Parts of such a charge can be thought
to propagate via the molecule from cathode to anode. Also, equal positive charges on both tips direct the hydrogen molecule axis mainly
transverse to the tip-tip line where it is uniformly distributed over the azimuthal angle. Then the distance between the tips and
the hydrogen electronic charge cloud, even if somewhat polarized and constricted in its lateral dimension, is larger than it is
in the case of no tip charge at all which corresponds to a spherically symmetric hydrogen charge cloud. And that additional distance will be still more
pronounced comparing with the case
of negatively charged tips with an elongated distribution towards the tips. Thus, changing the bias or the distance will open a channel
that, with suitable additional driving
voltage, may exchange electrons between the tips through the hydrogen bridge: less in the case of
positive tips and more in the case of negative tips.

In terms of surface physics, the validity of our calculations with respect to effects considered here is limited
to the field of physisorption, i.e. the H$_2$ molecule should not be dominated by covalent bonds to the tips. Including chemisorption
to the tips would not present a difficult computational obstacle, as one could add bonding orbitals centered on the tips. Starting at the  optimization level obtained here, one would need a similar number of additional computer runs with the advantage that the previous optimizations have
already paved the path towards the optimum, as the electrostatic environment can be composed from the basic dipole here considered to get a first guess for the optimization parameters.
A quantitative comparison with the exciting experiments
in the field of H$_2$ embedding in cavities~\cite{HuiWang},\cite{Carbonell} or H$_2$ bridges~\cite{Djukic} would need closer material
specifications and corresponding modelling to fit the scope of the presented QMC calculations. In any event, the striking resemblance of our
binding energy curves in Fig.~\ref{AFMvsdistanceandTC+} with the  theoretical curves related to the cavity experiments~\cite{HuiWang}
deserves closer consideration. Our results already show that the simple configuration of two point charges can mimic the trapping of the
H$_2$ molecule found in a cavity. The critical distance and the amount of binding energy differ by a factor of two which can be imagined
to be due to the details of the electrostatic field distributions and the caveat of the distance definition of realistic corrugated tips
to be modelled by point charges.

In terms of solid or liquid media, one could associate the negative or positive tip charge situation with an anionic or cationic
environment of a H$_2$ neutral molecule that acts differently on the probability distribution of its nuclear configuration, i.e. forcing
its direction parallel or perpendicular to the line connecting two opposing neighboring charges. It would need, besides the equal charge
configuration considered here, also charges of different signs, costing a similar computational amount but with the additional complication
that one would have to find the interesting and at least locally stable configurations among a large variety of configurations.

In summary, in a rather simple stable behavior, a hydrogen molecule
between two equally charged tips adjusts such as to let the protons
follow the external force, be it repulsion by positive charge or
attraction by negative charge, thereby incorporating a rotation of
the axis direction into a homogeneous distribution close to the
equatorial plane in the former case and along the poles in the latter
case. This picture should apply to more complicated
charge embedding of the molecule as well. The vibrational quantum eigenstates are complicated by the rotational polar angle degree of freedom coherently mixed into
radially quantized states after loosing
spherical symmetry by the tips. However, clear fingerprints of stretching modes are
observed as {\it breathing} modes in a spheroidal distribution of
the axis direction. We are not aware that such properties could have
been resolved in the BO approximation.
\section{Acknowledgements}
This work has been supported in part by the Basque Departamento de Educaci\'{o}n, Universidades e
Investigaci\'{o}n, the University of the Basque Country UPV/EHU (Grant No. IT1246-19) and the
Spanish Ministerio de Ciencia y Tecnolog\'{\i}a (Grant No. FIS2016-76471-P). The assistance of
the DIPC computer center is thankfully acknowledged. WS is indebted to the Institute of Theoretical
Physics and Astrophysics of CAU Kiel as well as to the university's computer center. MAVH acknowledges financial support from the Collaborative Research Fund of the Research Grants Council of Hong Kong (Grant No. C2014-15G). ICTS is supported by the HKBU Institute of Creativity, which is sponsored by the Hung Hin Shiu Charitable Foundation. 
\newpage
%
%
%
\newpage
\newpage
\newpage
\newpage
\newpage
%
%
%
\newpage
\newpage
\newpage
%
%
\newpage
%
\end{document}